\documentclass[twocolumn,notitlepage,tightenlines,preprintnumbers,amsmath,amssymb]{revtex4}

\usepackage{graphicx,amsmath}
\usepackage{epsf}
\usepackage{xcolor}

\begin{document}
\title{Photon and Phonon Spectral-Functions for Continuum Quantum Optomechanics}
\author{Hashem Zoubi}
\email{hashemz@hit.ac.il}
\affiliation{Department of Physics, Holon Institute of Technology, Holon 5810201, Israel}
\date{15 October, 2019}

\begin{abstract}
We study many-particle phenomena of propagating multi-mode photons and phonons interacting through Brillouin scattering type Hamiltonian in nanoscale waveguides. We derive photon and phonon retarded Green's functions and extract their spectral functions in applying the factorization approximation of the mean-field theory. The real part of the self-energy provides renormalization energy shifts for the photons and the phonons. Besides the conventional leaks, the imaginary part gives effective photon and phonon damping rates induced due to many-particle phenomena. The results extend the simple spectral functions of quantum optomechanics into continuum quantum optomechanics. We present the influence of thermal phonons on the photon effective damping rates, and consider cases of specific photon fields to be excited within the waveguide and which are of importance for phonon cooling scenarios.
\end{abstract}

\maketitle

\section{Introduction}

Controlling photons and phonons in nanoscale structures opened new horizons for photonics and phononics with applications, e.g. in quantum communication, quantum sensing and quantum information processing \cite{Eggleton2013,Safavi2019}. In nanoscale devices, for examples resonators and waveguides, photons and phonons are ideally hypridize within the same setup, where electromagnetic modes and mechanical vibrations spatially overlap \cite{Rakich2012,Sipe2016,Zoubi2016}. Optical photons of hundreds of nanometer wavelength match with that of gigahertz phonons propagating with sound velocity inside dielectric waveguides, e.g. made of silicon or silica. The high confinement of the electromagnetic field and of the mechanical waves in such nano structures lead to enhanced photon-phonon interactions (quantum Brillouin scattering), which are induced by conventional electrostriction and radiation presure \cite{Shin2013,Kittlaus2015,VanLaer2015a,VanLaer2015b,Huy2016}.

Extended nanoscale structures, with one dimensional confinement as for quantum wells or two dimensional confinement as for quantum wires, give rise to localized photon and phonon modes in the restricted direction, and they propagate freely in the extended direction. For example, a nanowire serves as a waveguide for both photons and phonons, and the combination of transverse confinement with extended structure yield photon and phonon branches \cite{Rakich2012,Zoubi2016}. Namely, the confinement provides discrete photon and phonon modes, while the extended structure with continuous translational symmetry allows photons and phonons to freely propagate with proper wavenumbers. The strong photon-phonon coupling in such extended nanostructures expands the study of conventional quantum optomechanics \cite{Aspelmeyer2014}, e.g. in resonators with discrete single modes, into continuum quantum optomechanics of propagating photon and phonon multi-modes \cite{Rakich2016}.

In the recent decade big progress have been achieved in the fabrication of nanowires (with dimension less than the optical wavelength and of about a centimeter length) \cite{Eggleton2013,Safavi2019}. The Brillouin gain have been shown to be several orders of magnitude larger than that of conventional fibers \cite{Shin2013,Kittlaus2015,VanLaer2015a,VanLaer2015b}. Coherent quantum phenomena in nanoscale waveguides have been studied experimentally, e.g. for silicon Brillouin laser \cite{Otterstrom2017} and photonic-phononic memory devices \cite{Zhu2007,Merklein2016}, and theoretically, e.g. for quantum logic gates \cite{Zoubi2017} and non-classical states \cite{Zoubi2018,Zoubi2019}. Inspite of the fact that photons show low damping rates, where waveguide photon leaks can be neglicted, phonon decoherence and losses are of big influence and put limitations on coherent performance of these components. Phonon dissipation in nanostructure are mainly due to geometric disorder, mechanical contacts and thermal noise \cite{Wuttke2013,Kharel2016,VanLaer2017}, and can strongly influence phonon cooling senarios \cite{Otterstrom2017}. Photon and phonon dissipations introduce a big challenge for the future of photonic and phononic efficiency in nanostructures.

In the present paper we study effective damping and frequency renormalization of coupled photons and phonons in one dimensional systems for continuum quantum optomechanics. We treat photons and phonons in nanoscale waveguides with Brillouin scattering type Hamiltonian, where the photons are scattered and the phonons emitted or absorbed, which are subjected to conservation of energy and momentum \cite{Zoubi2016}. We perform the calculations in the momentum space, where the photon and phonon states are specified by wavenumbers, which is typical for extended systems with transnational symmetry. Such many-particle system can be treated in using the known tool of Green's functions and in deriving their Spectral Functions (SF)s \cite{Kadanoff1962,Abrikosov1963,Fetter1971}. Moreover, we apply the factorization approximation in the mean field theoty \cite{Mahan2000}, in which photon and phonon correlations are negligible.

We extend the case of conventional quantum optomechanics \cite{Aspelmeyer2014}, that includes interactions among single photons and phonons, into continuum quantum optomechanics icluding multi-mode photons and phonons \cite{Rakich2016}. The use of conventional SFs, which were derived for single coupled photons and phonons \cite{Aspelmeyer2014,Clerk2010}, is inexact for a continuum setup with interacting multi-mode photons and phonons (for examole in \cite{Kharel2016,Wolff2017,VanLaer2017,Behunin2018}). Hence it is necessary to derive the proper continuum quantum optomechanical SFs in applying many-particle techniques. Besides the conventional photon and phonons damping rates, that are included phenomenologically in the present paper, we obtain effective damping rates for the photons and the phonons that appear due to many-particle phenomena. Here a photon scatters out of a specific mode into other modes, and other photon modes can be scatter into this specific mode, where the process involves the appearance or disappearance of phonons. On the other hand, a photon is accompanied with a cloud of phonons, which represents emission and absorption of virtual phonons, and that lead to a renormalization of the photon frequency. Furthermore, the phonons are accompanied with a cloud of virtual photons, which represents virtual scattering of photons, and that lead to a renormalization of the phonon frequency. We emphasize the effect of thermal phonons on the scattering of photons. We treat cases of specific photon modes to be excited with a fixed average number of photons within the waveguide, and discuss their influence on heating and cooling of phonons \cite{Bahl2012,Agarwal2013a,Otterstrom2018}.

The paper is opened in section 2 by presenting photon and phonon dispersions in nanoscale waveguides with emphasize on the lowest photon branch and the lowest two phonon branches. In section 3 we introduce phonon and photon Green's functions and SFs and give their effective damping rates and renormalization frequency shifts. We discuss in section 4 three cases of none, one and two photon fields to be excited within the waveguide with thermal equilibrium phonons.

\section{Interacting photons and phonons in nanoscale waveguides}

We consider a one dimensional nanoscale waveguide made of dielectric material that is embedded in free space. Nanoscale silicon waveguides have been realized for rectangular cross section wires with extension of hundreds of nanometers, e.g. $450\times750$~nm, with a length in the range from millimeters up to centimeters, and in applying light power with several milli-Watts and of wavelength $1550$~nm \cite{Safavi2019}. Other setups in use are tapered nanofibers made of silica where cylindrical waveguides of $500$~nm diameter and length of several centemeters have been achieved \cite{Vetsch2010}. In figure (1) a cylindrical waveguide is schematically  sketched.

Photons and phonons can propagate along the waveguide, but they are confined in its transverse direction. Namely, the transverse confinement gives rise to discrete modes, while for each mode the photons and phonon can propagate with a given dispersion, hence we obtain photon and phonon branches \cite{Zoubi2016}. The photons are described by the Hamiltonian
\begin{equation}
H_{phot}=\sum_{k\mu}\hbar\omega_{k\mu}\ a_{k\mu}^{\dagger}a_{k\mu},
\end{equation}
where $a_{k\mu}^{\dagger}$ and $a_{k\mu}$ are the creation and annihilation
operators of a photon of wavenumber $k$ at branch $\mu$. Here $\omega_{k\mu}$ is the photon angular frequency, where the wavenumber can be defined by $k=\frac{2\pi}{L}n$ with $(n=0,\pm1,\pm2,\cdots,\pm\infty)$, and $L$ is the effective waveguide length. The photon dispersions have been calculated numerically in \cite{Rakich2012} for rectangular cross section nanowire, and that agree with emperical results \cite{Eggleton2013,Safavi2019}. For cylindrical nanoscale waveguide we have calculated the dispersions analytically \cite{Zoubi2016}. Interestingly, at the lowest branch, within a significant interval of wavenumbers, a single photon mode can propagate with a linear dispersion $\omega_k=\omega_0+v_gk$, where $v_g$ is an effective group velocity, and a frequency $\omega_0$ appears due to the transverse confinement. For nanoscale waveguides we can take the typical physical numbers of $\omega_0/(2\pi)\approx 10^{14}$~Hz, with the light group velocity of $v_g\approx c/5$ (acording to our results for a nanofiber made of silicon \cite{Zoubi2016}). In figure (2) we plot the photon linear dispersion, $\omega_k$, as a function of $ka$, for the lowest photon branch. Here we consider a cylindrical waveguide of radius $a=250$~nm and with the length of one centimeter. Photons have relatively long lifetimes inside nanoscale waveguides \cite{Safavi2019}. The photon lifetimes can be considered by including damping rates, $\gamma_{k\mu}$, phenomenologically, where typical numbers for nanoscale waveguides are of about $10^4-10^5$~Hz, which can be extracted from experimental observations.

\begin{figure}
\includegraphics[width=0.8\linewidth]{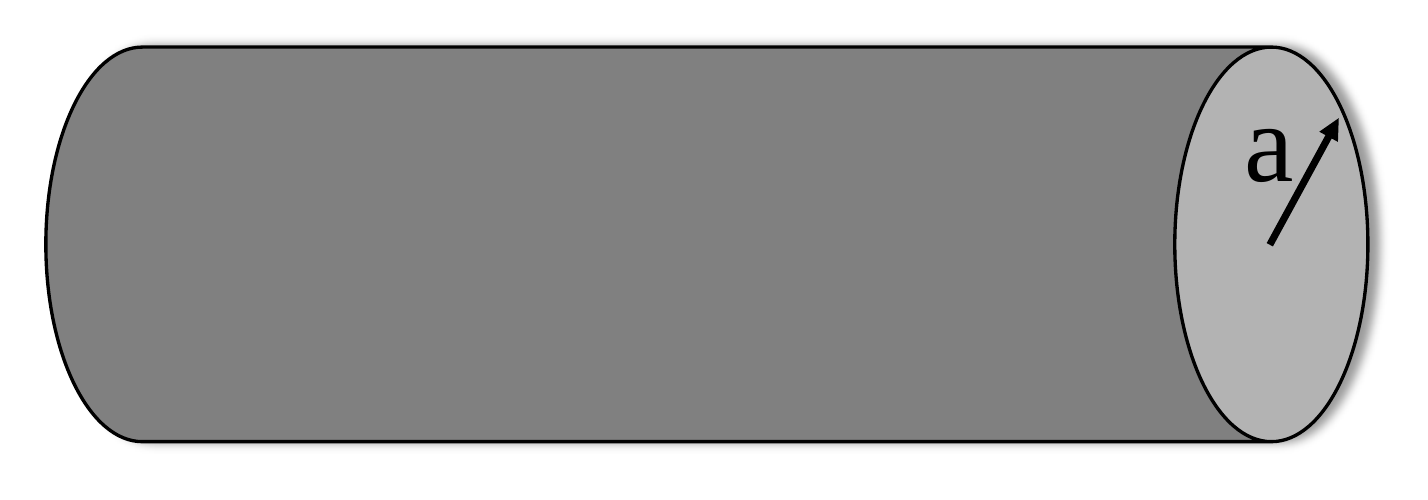}
\caption{Nanoscale waveguide of circular cross section with radios $a$, made, e.g., of silicon or silica and embedded in free space. The length can be of several centimeters.}
\label{PhotPhonDis}
\end{figure}

\begin{figure}
\includegraphics[width=0.8\linewidth]{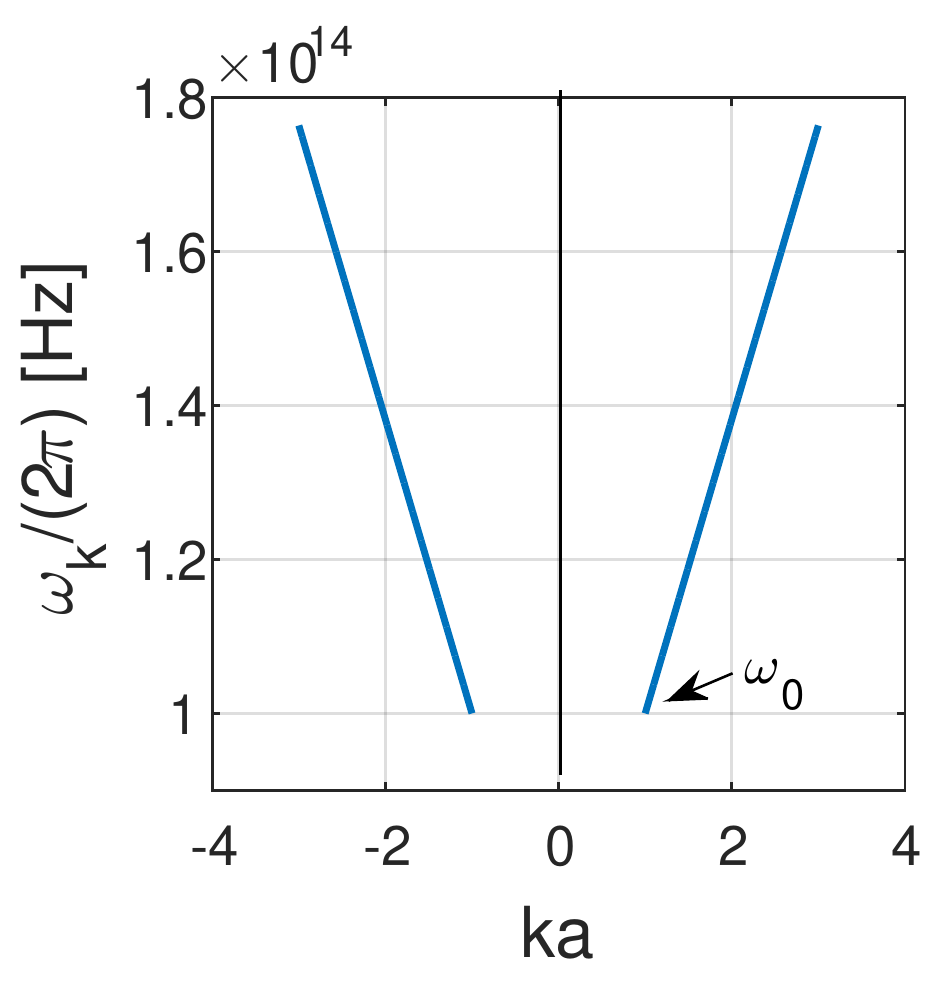}
\caption{The photon dispersion $\omega_k=\omega_0+v_gk$ vs. $ka$ for the lowest branch in the linear zone. Here $\omega_0/(2\pi)\approx 10^{14}$~Hz, with $a=250$~nm, and the group velocity is $v_g\approx c/5$.}
\label{PhotPhonDis}
\end{figure}

On the other side, the phonons are described by the Hamiltonian
\begin{equation}
H_{phon}=\sum_{q\alpha}\hbar\Omega_{q\alpha}\ b_{q\alpha}^{\dagger}b_{q\alpha},
\end{equation}
where $b_{q\alpha}^{\dagger}$ and $b_{q\alpha}$ are the creation and annihilation
operators of a phonon of wavenumber $q$ at branch $\alpha$ of angular frequency
$\Omega_{q\alpha}$. Several phonon branches can be excited inside nanoscale waveguides. The lowest branch is an acoustic mode with linear dispersion, where $\Omega_q=v_aq$ with the sound velocity $v_a$. Among the higher branches are dispersionless vibrational modes with frequency $\Omega_v$. Several lowest mechanical excitation branches have been calculated numerically for rectangular cross section waveguides made of silicon \cite{Rakich2012}, and which are confirmed experimentally \cite{Eggleton2013,Safavi2019}. In our previous work \cite{Zoubi2016} we calculated the lowest phonon branches for cylindrical nanofibers made of silicon. For example, in silicon cylindrical nanoscale waveguide, the lowest
phonon branch is of sound wave with velocity $v_a=8433$~m/s, and the second excited branch is of
vibrational mode of frequency $\Omega_v/(2\pi)\approx10$~GHz \cite{Zoubi2016}. In figure (3), we present the phonon lowest acoustic branch and the lowest vibrational mode, where we plot the phonon dispersion, $\Omega_q$, as a function of $qa$. Phonon lifetimes of mechanical excitations are limited by structural and thermal disorders \cite{Safavi2019}, and can be included phenomenologically through the damping rates, $\Gamma_{q\alpha}$. Typical numbers for phonon damping rates in silicon nanowires are in the range of $1-10$~MHz, as observed in several experiments \cite{Safavi2019}.

\begin{figure}
\includegraphics[width=0.8\linewidth]{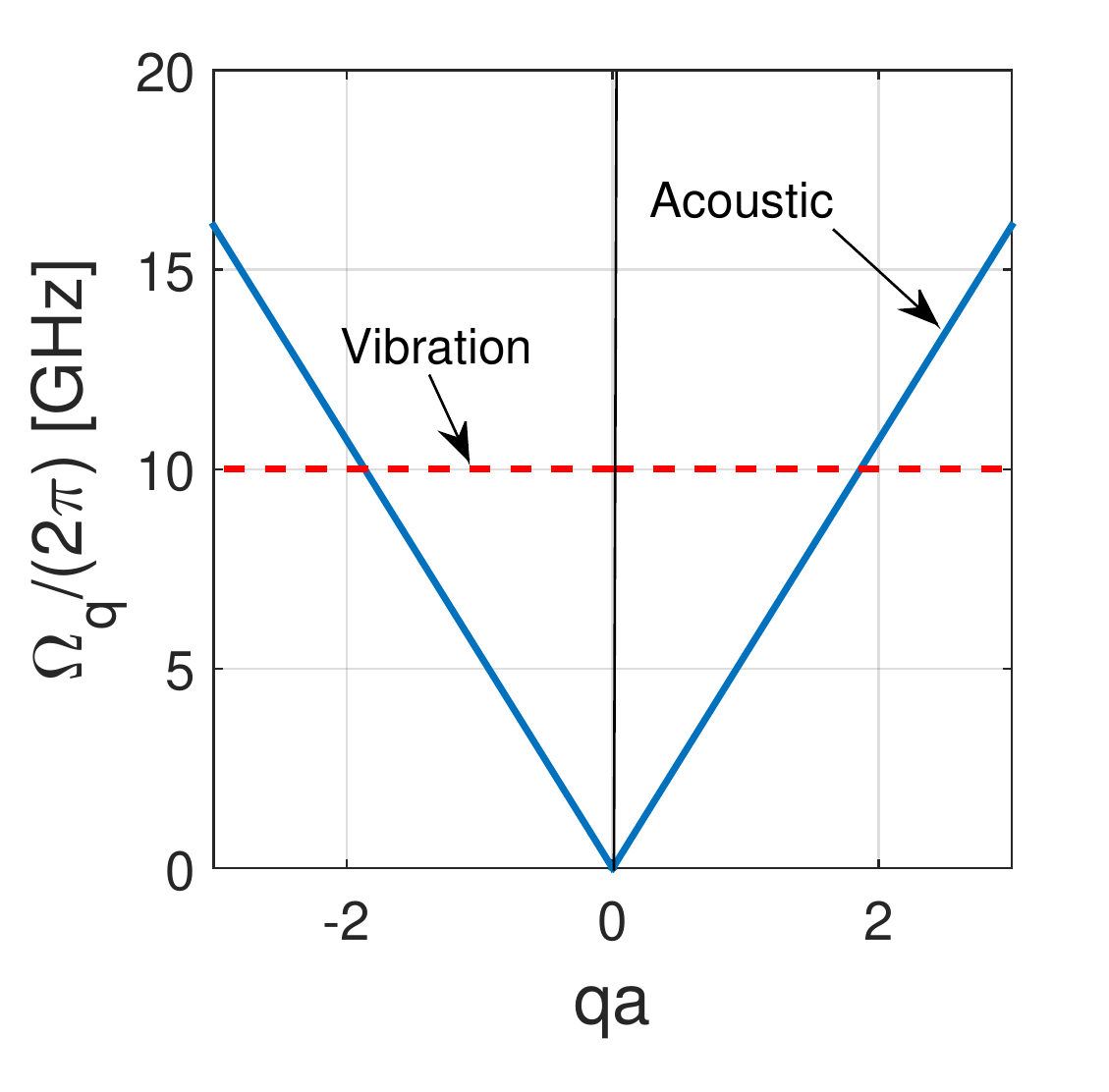}
\caption{The acoustic phonon dispersion, $\Omega_q=v_aq$ vs. $qa$. Here, the sound velocity in silicon is $v_a=8433$~m/s. The horizontal dashed line is for the vibrational mode of frequency $\Omega_v/(2\pi)\approx10$~GHz.}
\label{PhotPhonDis}
\end{figure}

Inside nanoscale waveguides the photons and phonons are strongly interact due to electrostriction and radiation pressure. In our previous work we studied the system of coupled photons and phonons inside nanoscale waveguides in deriving the microscopic Brillouin scattering type Hamiltonian \cite{Zoubi2016}. The photon-phonon interaction is given by
\begin{equation}
H_{phot-phon}=\sum_{k\mu}\sum_{q\alpha}f_{k\mu}^{q\alpha}\ a_{k+q\mu}^{\dagger}a_{k\mu}B_{q\alpha},
\end{equation}
where $f_{k\mu}^{q\alpha}$ is the photon-phonon coupling parameter, and we defined the operator
\begin{equation}
B_{q\alpha}=b_{q\alpha}+b_{-q\alpha}^{\dagger}.
\end{equation}
Note that $B_{q\alpha}=B^{\dagger}_{-q\alpha}$. Detail calculations of $f_{k\mu}^{q\alpha}$ appear in \cite{Zoubi2016}.

We have been analytically calculated the photon-phonon coupling parameters for a cylindrical nanoscale waveguide made of silicon in considering electrostriction and radiation pressure \cite{Zoubi2016}. Moreover, in \cite{Rakich2012} the calculation have been done numerically for silicon rectangular waveguides. The coupling parameters can be extracted experimentally from the observed Brillouin gain parameters and found to be in agreement with the calculated ones \cite{Shin2013,Kittlaus2015,VanLaer2015a,VanLaer2015b}, where the coupling parameters are mainly in the range of $1-10$~MHz. The strong photon-phonon interactions are achieved due to the large overlaps between mechanical excitations and electromagnetic fields inside the nanowires. This fact is clear from the similarity of the photon and phonon wavelngths inside the waveguide. For $10$~GHz phonons one gets $1\ \mu$m wavelength for $10^4$~m/s sound velocity, and for $10^{14}$~Hz photons one get $1\ \mu$m wavelength for $10^8$~m/s effective light velocity \cite{Safavi2019}.

\section{Photon and Phonon Spectral-Functions}

Green's functions are an important tool for achieving physical properties for a system of many-particle interacting photons and phonons \cite{Kharel2016,Wolff2017,VanLaer2017,Behunin2018}. Physical properties can be extracted from Green's functions, which provide renormalized energies and lifetimes of quasi-particles. We concentrate here in the observable Spectral-Functions (SF)s of photons and phonons that are defined as minus twice the imaginary part of the photon and phonon retarded Green's function according to the fluctuaction-dissipation theorem \cite{Kadanoff1962,Abrikosov1963,Fetter1971}. Namely, we have
\begin{eqnarray} \label{SF}
{\cal S}_{k\mu}^{phot}(\omega)&=&-2\ \Im\left\{{\cal
  G}_{k\mu}^{phot}(\omega)\right\},\nonumber \\
{\cal S}_{q\alpha}^{phon}(\omega)&=&-2\ \Im\left\{{\cal
  D}_{q\alpha}^{phon}(\omega)\right\}.
\end{eqnarray}
The retarded Green's functions ${\cal G}_{k\mu}^{phot}(\omega)$ and ${\cal D}_{q\alpha}^{phon}(\omega)$ are the Fourier transform, that is
\begin{equation}\label{Fourier}
{\cal F}(t)=\frac{1}{2\pi}\int_{-\infty}^{+\infty}d\omega\ e^{-i\omega t}{\cal
  F}(\omega),
\end{equation}
of the retarded Green's functions \cite{Kadanoff1962,Abrikosov1963,Fetter1971,Mahan2000}
\begin{eqnarray}
{\cal G}_{k\mu}^{phot}(t)&=&-i\Theta(t)\left\langle\left[a_{k\mu}(t),a_{k\mu}^{\dagger}(0)\right]\right\rangle, \nonumber \\
{\cal D}_{q\alpha}^{phon}(t)&=&-i\Theta(t)\left\langle\left[b_{q\alpha}(t),b_{q\alpha}^{\dagger}(0)\right]\right\rangle,
\end{eqnarray}
where the step function is given by $\Theta(t)=1$ for $t>0$ and $\Theta(t)=0$ for $t<0$.

In the following we present the detail derivations of the Green's functions and SFs for multimodes of interacting photons and phonons. The Green's function equations of motion lead to a hierarchy of equations that can be truncated in applying known factorization approximation of mean-field theory, and furthermore in assuming time-independent average number of photons and phonons inside the waveguide \cite{Mahan2000}. Moreover, we include phenomenologically the photon and phonon damping rates, $\gamma_{k\mu}$ and $\Gamma_{q\alpha}$, in using the replacement $\omega_{k\mu}\rightarrow\omega_{k\mu}-i\gamma_{k\mu}/2$ and $\Omega_{q\alpha}\rightarrow\Omega_{q\alpha}-i\Gamma_{q\alpha}/2$.

\subsection{Green's Function Equations of Motion}

We start by deriving the photon and phonon operator equations of motion, which are
\begin{eqnarray}
i\frac{d}{dt}a_{k\mu}&=&\omega_{k\mu}\ a_{k\mu}+\sum_{q'\alpha'}f_{k+q'\mu}^{-q'\alpha'}\ a_{k+q'\mu}B_{q'\alpha'}^{\dagger},
\nonumber \\
i\frac{d}{dt}b_{q\alpha}&=&\Omega_{q\alpha}\ b_{q\alpha}+\sum_{k'\mu'}f_{k'\mu'}^{-q\alpha}\ a_{k'-q\mu'}^{\dagger}a_{k'\mu'}.\end{eqnarray}
Note that we have the symmetry relation $f_{k+q\mu}^{-q\alpha}=f_{k\mu}^{q\alpha}$, that appears due to symmetry in the photon and phonon dispersions.

The photon and phonon Green's function equations of motion are
\begin{eqnarray} \label{GFEs}
i\frac{d}{dt}{\cal G}_{k\mu}^{phot}(t)&=&\delta(t)+\omega_{k\mu}\ {\cal
  G}_{k\mu}^{phot}(t)\nonumber \\
&+&\sum_{q\alpha}f_{k\mu}^{q\alpha}\left\{{\cal P}_1(t)+{\cal
P}_2(t)\right\}, \nonumber \\
i\frac{d}{dt}{\cal D}_{q\alpha}^{phon}(t)&=&\delta(t)+\Omega_{q\alpha}\ {\cal
  D}_{q\alpha}^{phon}(t)\nonumber \\
&+&\sum_{k\mu}f_{k\mu}^{-q\alpha}\ {\cal P}_3(t),
\end{eqnarray}
using $\frac{d}{dt}\Theta(t)=\delta(t)$, where we defined
\begin{eqnarray}
{\cal P}_1(t)&=&-i\Theta(t)\left\langle\left[a_{k+q\mu}(t)b_{q\alpha}^{\dagger}(t),a_{k\mu}^{\dagger}(0)\right]\right\rangle, \nonumber \\
{\cal P}_2(t)&=&-i\Theta(t)\left\langle\left[a_{k+q\mu}(t)b_{-q\alpha}(t),a_{k\mu}^{\dagger}(0)\right]\right\rangle, \nonumber \\
{\cal
  P}_3(t)&=&-i\Theta(t)\left\langle\left[a_{k-q\mu}^{\dagger}(t)a_{k\mu}(t),b_{q\alpha}^{\dagger}(0)\right]\right\rangle.
\end{eqnarray}
We treat now separately the functions ${\cal P}_i(t)$ with $(i=1,2,3)$.

\subsubsection{The ${\cal P}_1(t)$ function}

The ${\cal P}_1(t)$ function equation of motion is
\begin{eqnarray}
i\frac{d}{dt}{\cal P}_1(t)&=&\delta(t)\delta_{q,0}\left\langle
b_{0\alpha}^{\dagger}(t)\right\rangle+(\omega_{k+q\mu}-\Omega_{q\alpha}^{\ast})\ {\cal P}_1(t)
\nonumber \\
&-&i\Theta(t)\sum_{q'\alpha'}f_{k+q+q'\mu}^{-q'\alpha'}\nonumber \\
&\times &\left\langle\left[a_{k+q+q'\mu}(t)B_{q'\alpha'}^{\dagger}(t)b_{q\alpha}^{\dagger}(t),a_{k\mu}^{\dagger}(0)\right]\right\rangle
\nonumber \\
&+&i\Theta(t)\sum_{k'\mu'}f_{k'\mu'}^{q\alpha}\nonumber \\
&\times &\left\langle\left[a_{k+q\mu}(t)a_{k'+q\mu'}^{\dagger}(t)a_{k'\mu'}(t),a_{k\mu}^{\dagger}(0)\right]\right\rangle.\nonumber \\
\end{eqnarray}
We neglect the first term, by assuming $\left\langle
b_{0\alpha}^{\dagger}(t)\right\rangle\approx 0$. The last two terms of the right side are problematic as their equations of motion lead to higher order terms. This result leads to infinite hierarchy of equations. We overcome this difficulty by appeal to the factorization approximation of the mean field theory \cite{Kadanoff1962,Abrikosov1963,Fetter1971,Mahan2000}. We neglect photon-phonon correlations, as coherence between them decay very fast. Hence, in the mean-field theory we factorize the expectation values into a product of photon and phonon terms. We apply now the factorization approximation in taking
\begin{eqnarray}
&&\left\langle\left[a_{k+q+q'\mu}(t)B_{q'\alpha'}^{\dagger}(t)b_{q\alpha}^{\dagger}(t),a_{k\mu}^{\dagger}(0)\right]\right\rangle\nonumber \\
&\approx& \left\langle
B_{q'\alpha'}^{\dagger}(t)b_{q\alpha}^{\dagger}(t)\right\rangle\left\langle
\left[a_{k+q+q'\mu}(t),a_{k\mu}^{\dagger}(0)\right]\right\rangle, \nonumber \\
&&\left\langle\left[a_{k+q\mu}(t)a_{k'+q\mu'}^{\dagger}(t)a_{k'\mu'}(t),a_{k\mu}^{\dagger}(0)\right]\right\rangle\nonumber \\
&\approx&  \left\langle a_{k+q\mu}(t)a_{k'+q\mu'}^{\dagger}(t)\right\rangle\left\langle\left[a_{k'\mu'}(t),a_{k\mu}^{\dagger}(0)\right]\right\rangle.
\end{eqnarray}
We have
\begin{eqnarray}
\left\langle B_{q'\alpha'}^{\dagger}(t)b_{q\alpha}^{\dagger}(t)\right\rangle
&=&\left\langle b_{-q'\alpha'}(t)b_{q\alpha}^{\dagger}(t)\right\rangle \nonumber \\
&=&
\delta_{\alpha,\alpha'}\delta_{q,-q'}\left(1+n_{q\alpha}\right), \nonumber \\
 \left\langle a_{k+q\mu}(t)a_{k'+q\mu'}^{\dagger}(t)\right\rangle &=&
 \delta_{k,k'}\delta_{\mu,\mu'} \left(1+N_{k+q\mu}\right),
\end{eqnarray}
where $n_{q\alpha}$ is the average number of phonons at mode $(q\alpha)$, and $N_{k\mu}$ is the average number of photons at mode $(k\mu)$, which are defined by
\begin{equation}\label{SSNn}
N_{k\mu}=\left\langle a_{k\mu}^{\dagger}a_{k\mu}\right\rangle,\ n_{q\alpha}=\left\langle b_{q\alpha}^{\dagger}b_{q\alpha}\right\rangle.
\end{equation}
Moreover, we assumed that the average number of photons and phonons inside the waveguide is in steady state, which is possible for the photons by a combination of external pump fields and dissipation, and for thermal equilibrium phonons. The approximation gives the result
\begin{eqnarray}\label{P1E}
i\frac{d}{dt}{\cal P}_1(t)&=&(\omega_{k+q\mu}-\Omega_{q\alpha}^{\ast})\ {\cal P}_1(t)
\nonumber \\
&+&
f_{k\mu}^{q\alpha}\left\{\left(1+n_{q\alpha}\right)-\left(1+N_{k+q\mu}\right)\right\}\ {\cal
  G}_{k\mu}^{phot}(t).\nonumber \\
\end{eqnarray}

\subsubsection{The ${\cal P}_2(t)$ function}

Next, the ${\cal P}_2(t)$ function equation of motion is
\begin{eqnarray}
i\frac{d}{dt}{\cal P}_2(t)&=&\delta(t)\delta_{q,0}\left\langle
b_{0\alpha}(t)\right\rangle+(\omega_{k+q\mu}+\Omega_{q\alpha})\ {\cal P}_2(t)
\nonumber \\
&-&i\Theta(t)\sum_{q'\alpha'}f_{k+q+q'\mu}^{-q'\alpha'}\nonumber \\
&\times &\left\langle\left[a_{k+q+q'\mu}(t)B_{-q'\alpha'}(t)b_{-q\alpha}(t),a_{k\mu}^{\dagger}(0)\right]\right\rangle
\nonumber \\
&-&i\Theta(t)\sum_{k'\mu'}f_{k'\mu'}^{q\alpha}\nonumber \\
&\times &\left\langle\left[a_{k+q\mu}(t)a_{k'+q\mu'}^{\dagger}(t)a_{k'\mu'}(t),a_{k\mu}^{\dagger}(0)\right]\right\rangle,\nonumber \\
\end{eqnarray}
We neglect the first term, where we assume $\left\langle
b_{0\alpha}(t)\right\rangle\approx 0$. As before, we apply the factorization approximation in taking
\begin{eqnarray}
&&\left\langle\left[a_{k+q+q'\mu}(t)B_{-q'\alpha'}(t)b_{-q\alpha}(t),a_{k\mu}^{\dagger}(0)\right]\right\rangle\nonumber \\
&\approx& \left\langle
B_{q'\alpha'}^{\dagger}(t)b_{-q\alpha}(t)\right\rangle \left\langle\left[a_{k+q+q'\mu}(t),a_{k\mu}^{\dagger}(0)\right]\right\rangle,
\nonumber \\
&&\left\langle\left[a_{k+q\mu}(t)a_{k'+q\mu'}^{\dagger}(t)a_{k'\mu'}(t),a_{k\mu}^{\dagger}(0)\right]\right\rangle\nonumber \\
&\approx& \left\langle a_{k+q\mu}(t)a_{k'+q\mu'}^{\dagger}(t)\right\rangle \left\langle\left[a_{k'\mu'}(t),a_{k\mu}^{\dagger}(0)\right]\right\rangle.
\end{eqnarray}
We have
\begin{eqnarray}
\left\langle B_{q'\alpha'}^{\dagger}(t)b_{-q\alpha}(t)\right\rangle &=&
\left\langle b_{q'\alpha'}^{\dagger}(t)b_{-q\alpha}(t)\right\rangle \nonumber \\
&=&
\delta_{\alpha,\alpha'}\delta_{-q,q'}\ n_{q\alpha},\nonumber \\
\left\langle a_{k+q\mu}(t)a_{k'+q\mu'}^{\dagger}(t)\right\rangle &=& \delta_{k,k'}\delta_{\mu,\mu'} \left(1+N_{k+q\mu}\right).
\end{eqnarray}
Finally, we get the result
\begin{eqnarray}\label{P2E}
i\frac{d}{dt}{\cal P}_2(t)&=&(\omega_{k+q\mu}+\Omega_{q\alpha})\ {\cal P}_2(t)
\nonumber \\
&+&
f_{k\mu}^{q\alpha}\left\{n_{q\alpha}+\left(1+N_{k+q\mu}\right)\right\}\ {\cal
  G}_{k\mu}^{phot}(t).
\end{eqnarray}

\subsubsection{The ${\cal P}_3(t)$ function}

The ${\cal P}_3(t)$ function equation of motion is
\begin{eqnarray}
i\frac{d}{dt}{\cal P}_3(t)&=&(\omega_{k\mu}-\omega_{k-q\mu}^{\ast})\ {\cal P}_3(t)
\nonumber \\
&+&i\Theta(t)\sum_{q'\alpha'}f_{k-q+q'\mu}^{-q'\alpha'}\nonumber \\
&\times &\left\langle\left[a_{k-q+q'\mu}^{\dagger}(t)B_{q'\alpha'}(t)a_{k\mu}(t),b_{q\alpha}^{\dagger}(0)\right]\right\rangle
\nonumber \\
&-&i\Theta(t)\sum_{q'\alpha'}f_{k+q'\mu}^{-q'\alpha'}\nonumber \\
&\times &\left\langle\left[a_{k-q\mu}^{\dagger}(t)a_{k+q'\mu}(t)B_{q'\alpha'}^{\dagger}(t),b_{q\alpha}^{\dagger}(0)\right]\right\rangle.\nonumber \\
\end{eqnarray}
Once more, we apply the factorization approximation in taking
\begin{eqnarray}
&&\left\langle\left[a_{k-q+q'\mu}^{\dagger}(t)B_{q'\alpha'}(t)a_{k\mu}(t),b_{q\alpha}^{\dagger}(0)\right]\right\rangle\nonumber \\
&\approx& \left\langle
a_{k-q+q'\mu}^{\dagger}(t)a_{k\mu}(t)\right\rangle\left\langle\left[B_{q'\alpha'}(t),b_{q\alpha}^{\dagger}(0)\right]\right\rangle
\nonumber \\
&&\left\langle\left[a_{k-q\mu}^{\dagger}(t)a_{k+q'\mu}(t)B_{q'\alpha'}^{\dagger}(t),b_{q\alpha}^{\dagger}(0)\right]\right\rangle\nonumber \\
&\approx& \left\langle a_{k-q\mu}^{\dagger}(t)a_{k+q'\mu}(t)\right\rangle \left\langle\left[B_{q'\alpha'}^{\dagger}(t),b_{q\alpha}^{\dagger}(0)\right]\right\rangle.
\end{eqnarray}
Using
\begin{eqnarray}
\left\langle a_{k-q+q'\mu}^{\dagger}(t)a_{k\mu}(t)\right\rangle &=&
\delta_{q,q'}N_{k\mu} \nonumber \\
\left\langle a_{k-q\mu}^{\dagger}(t)a_{k+q'\mu}(t)\right\rangle &=& \delta_{-q,q'}N_{k-q\mu},
\end{eqnarray}
we get
\begin{eqnarray}
&&\left\langle\left[a_{k-q+q'\mu}^{\dagger}(t)B_{q'\alpha'}(t)a_{k\mu}(t),b_{q\alpha}^{\dagger}(0)\right]\right\rangle\nonumber \\
&\approx& \delta_{q,q'}\delta_{\alpha,\alpha'}N_{k\mu} \left\langle\left[b_{q'\alpha'}(t),b_{q\alpha}^{\dagger}(0)\right]\right\rangle
\nonumber \\
&&\left\langle\left[a_{k-q\mu}^{\dagger}(t)a_{k+q'\mu}(t)B_{q'\alpha'}^{\dagger}(t),b_{q\alpha}^{\dagger}(0)\right]\right\rangle\nonumber \\
&\approx& \delta_{-q,q'}\delta_{\alpha,\alpha'}N_{k-q\mu} \left\langle\left[b_{-q'\alpha'}(t),b_{q\alpha}^{\dagger}(0)\right]\right\rangle.
\end{eqnarray}
Finally we have
\begin{eqnarray}\label{P3E}
i\frac{d}{dt}{\cal P}_3(t)&=&(\omega_{k\mu}-\omega_{k-q\mu}^{\ast})\ {\cal P}_3(t)
\nonumber \\
&+&f_{k-q\mu}^{q\alpha}\left\{N_{k-q\mu}- N_{k\mu}\right\}\ {\cal D}_{q\alpha}^{phon}(t).
\end{eqnarray}

\subsubsection{Green's Functions in Fourier Space}

In applying the Fourier Transform of Eq.(\ref{Fourier}), and using the relation
\begin{equation}
\delta(t)=\frac{1}{2\pi}\int_{-\infty}^{+\infty}d\omega\ e^{-i\omega t},
\end{equation}
we get algebraic system of equations. For photon and photon Green's functions, using equations (\ref{GFEs}), we get
\begin{eqnarray}
\omega\ {\cal D}_{q\alpha}^{phon}(\omega)&=&1+\Omega_{q\alpha}\ {\cal
  D}_{q\alpha}^{phon}(\omega)+\sum_{k\mu}f_{k\mu}^{-q\alpha}\ {\cal P}_3(\omega), \nonumber \\
\omega\ {\cal G}_{k\mu}^{phot}(\omega)&=&1+\omega_{k\mu}\ {\cal
  G}_{k\mu}^{phot}(\omega)\nonumber \\
&+&\sum_{q\alpha}f_{k\mu}^{q\alpha}\left\{{\cal P}_1(\omega)+{\cal
P}_2(\omega)\right\},
\end{eqnarray}
and for the ${\cal P}_i$ functions, using equations (\ref{P1E},\ref{P2E},\ref{P3E}), we get
\begin{eqnarray}
\omega\ {\cal P}_1(\omega)&=&(\omega_{k+q\mu}-\Omega_{q\alpha}^{\ast})\ {\cal P}_1(\omega)\nonumber \\
&+&f_{k\mu}^{q\alpha}\left\{\left(1+n_{q\alpha}\right)-\left(1+N_{k+q\mu}\right)\right\}\ {\cal
  G}_{k\mu}^{phot}(\omega), \nonumber \\
\omega\ {\cal P}_2(\omega)&=&(\omega_{k+q\mu}+\Omega_{q\alpha})\ {\cal P}_2(\omega)\nonumber \\
&+&f_{k\mu}^{q\alpha}\left\{n_{q\alpha}+\left(1+N_{k+q\mu}\right)\right\}\ {\cal
  G}_{k\mu}^{phot}(\omega), \nonumber \\
\omega\ {\cal P}_3(\omega)&=&(\omega_{k\mu}-\omega_{k-q\mu}^{\ast})\ {\cal P}_3(\omega)\nonumber \\
&+&f_{k-q\mu}^{q\alpha}\left\{N_{k-q\mu}- N_{k\mu}\right\}\ {\cal
  D}_{q\alpha}^{phon}(\omega).
\end{eqnarray}
Solving for the ${\cal P}_i(\omega)$ functions in term of the Green functions ${\cal G}_{k\mu}^{phot}(\omega)$ and ${\cal D}_{q\alpha}^{phon}(\omega)$, we get
\begin{eqnarray}
{\cal P}_1(\omega)&=&f_{k\mu}^{q\alpha}\ \frac{n_{q\alpha}-N_{k+q\mu}}{\omega-\omega_{k+q\mu}+\Omega_{q\alpha}^{\ast}}\ {\cal
  G}_{k\mu}^{phot}(\omega), \nonumber \\
{\cal P}_2(\omega)&=&f_{k\mu}^{q\alpha}\ \frac{1+n_{q\alpha}+N_{k+q\mu}}{\omega-\omega_{k+q\mu}-\Omega_{q\alpha}}\ {\cal
  G}_{k\mu}^{phot}(\omega), \nonumber \\
{\cal P}_3(\omega)&=&f_{k-q\mu}^{q\alpha}\ \frac{N_{k-q\mu}- N_{k\mu}}{\omega-\omega_{k\mu}+\omega_{k-q\mu}^{\ast}}\ {\cal
  D}_{q\alpha}^{phon}(\omega).
\end{eqnarray}
Using these results in the equations for the Green functions, yields
\begin{eqnarray}
{\cal G}_{k\mu}^{phot}(\omega)&=&\frac{1}{\omega-\omega_{k\mu}-{\cal M}_{k\mu}^{phot}(\omega)},
\nonumber \\
{\cal D}_{q\alpha}^{phon}(\omega)&=&\frac{1}{\omega-\Omega_{q\alpha}-{\cal M}_{q\alpha}^{phon}(\omega)},
\end{eqnarray}
where the self-energy functions are given by
\begin{eqnarray}
{\cal M}_{k\mu}^{phot}(\omega)&=&\sum_{q\alpha}\left(f_{k\mu}^{q\alpha}\right)^2\left\{\frac{n_{q\alpha}-N_{k+q\mu}}{\omega-\omega_{k+q\mu}+\Omega_{q\alpha}^{\ast}}\right.\nonumber \\
&+&\left.\frac{1+n_{q\alpha}+N_{k+q\mu}}{\omega-\omega_{k+q\mu}-\Omega_{q\alpha}}\right\},
\nonumber \\
{\cal M}_{q\alpha}^{phon}(\omega)&=&\sum_{k\mu}\left(f_{k\mu}^{-q\alpha}\right)^2\ \frac{N_{k-q\mu}- N_{k\mu}}{\omega-\omega_{k\mu}+\omega_{k-q\mu}^{\ast}},
\end{eqnarray}
which are Dyson's type equations.

\subsection{Spectral Functions}

The photon and phonon self energy functions can be written in terms of real and imaginary parts. In including the photon and phonon conventional damping rates explicitly, by using the replacement $\omega_{k\mu}\rightarrow\omega_{k\mu}-i\gamma_{k\mu}/2$ and $\Omega_{q\alpha}\rightarrow\Omega_{q\alpha}-i\Gamma_{q\alpha}/2$, we have
\begin{eqnarray}
{\cal M}_{k\mu}^{phot}(\omega)&=&\Delta_{k\mu}^{phot}(\omega)-i\Lambda_{k\mu}^{phot}(\omega)/2,
\nonumber \\
{\cal M}_{q\alpha}^{phon}(\omega)&=&\Delta_{q\alpha}^{phon}(\omega)-i\Lambda_{q\alpha}^{phon}(\omega)/2.
\end{eqnarray}
Hence, the photon and phonon SFs, using Eqs.(\ref{SF}), are given by
\begin{eqnarray} \label{SFPP}
&&{\cal
  S}_{k\mu}^{phot}(\omega)=\nonumber \\
&&\frac{\gamma_{k\mu}+\Lambda_{k\mu}^{phot}(\omega)}{\left(\omega-\omega_{k\mu}-\Delta_{k\mu}^{phot}(\omega)\right)^2+\left(\gamma_{k\mu}+\Lambda_{k\mu}^{phot}(\omega)\right)^2/4},\nonumber
\\
&&{\cal S}_{q\alpha}^{phon}(\omega)=\nonumber \\
&&\frac{\Gamma_{q\alpha}+\Lambda_{q\alpha}^{phon}(\omega)}{\left(\omega-\Omega_{q\alpha}-\Delta_{q\alpha}^{phon}(\omega)\right)^2+\left(\Gamma_{q\alpha}+\Lambda_{q\alpha}^{phon}(\omega)\right)^2/4}.\nonumber \\
\end{eqnarray}
The phonon renormalization frequency shifts, $\Delta_{q\alpha}^{phon}(\omega)$,
and effective broadening, $\Lambda_{q\alpha}^{phon}(\omega)$, are induced by the existence of the photons. The photon renormalization frequency shift, $\Delta_{k\mu}^{phot}(\omega)$,
and effective broadening, $\Lambda_{k\mu}^{phot}(\omega)$, contain two parts that induced due to phonons and the appearance of other photon modes,
where we can write $\Delta_{k\mu}^{phot}(\omega)=\Delta_{k\mu}^{M}(\omega)+\Delta_{k\mu}^{EM}(\omega)$ and
$\Lambda_{k\mu}^{phot}(\omega)=\Lambda_{k\mu}^{M}(\omega)+\Lambda_{k\mu}^{EM}(\omega)$. The effect of the photons gives
\begin{eqnarray}
\Delta_{k\mu}^{EM}(\omega)&=&\sum_{q\alpha}\left(f_{k\mu}^{q\alpha}\right)^2N_{k+q\mu}\nonumber \\
&\times &\left\{\frac{\omega-\omega_{k+q\mu}-\Omega_{q\alpha}}{\left(\omega-\omega_{k+q\mu}-\Omega_{q\alpha}\right)^2+\left(\Gamma_{q\alpha}+\gamma_{k+q\mu}\right)^2/4}\right. \nonumber
\\
&-&
\left.\frac{\omega-\omega_{k+q\mu}+\Omega_{q\alpha}}{\left(\omega-\omega_{k+q\mu}+\Omega_{q\alpha}\right)^2+\left(\Gamma_{q\alpha}+\gamma_{k+q\mu}\right)^2/4}\right\}, \nonumber
\\
\Lambda_{k\mu}^{EM}(\omega)&=&\sum_{q\alpha}\left(f_{k\mu}^{q\alpha}\right)^2N_{k+q\mu}\nonumber \\
&\times &\left\{\frac{\Gamma_{q\alpha}+\gamma_{k+q\mu}}{\left(\omega-\omega_{k+q\mu}-\Omega_{q\alpha}\right)^2+\left(\Gamma_{q\alpha}+\gamma_{k+q\mu}\right)^2/4}\right. \nonumber
\\
&-&
\left.\frac{\Gamma_{q\alpha}+\gamma_{k+q\mu}}{\left(\omega-\omega_{k+q\mu}+\Omega_{q\alpha}\right)^2+\left(\Gamma_{q\alpha}+\gamma_{k+q\mu}\right)^2/4}\right\}.\nonumber \\
\end{eqnarray}
and the effect of the phonons yields
\begin{eqnarray} \label{DLM}
&&\Delta_{k\mu}^{M}(\omega)=\sum_{q\alpha}\left(f_{k\mu}^{q\alpha}\right)^2 \nonumber \\
&&\times \left\{n_{q\alpha}\ \frac{\omega-\omega_{k+q\mu}+\Omega_{q\alpha}}{\left(\omega-\omega_{k+q\mu}+\Omega_{q\alpha}\right)^2+\left(\Gamma_{q\alpha}+\gamma_{k+q\mu}\right)^2/4}\right. \nonumber
\\
&&+ \left.\left(1+n_{q\alpha}\right)\frac{\omega-\omega_{k+q\mu}-\Omega_{q\alpha}}{\left(\omega-\omega_{k+q\mu}-\Omega_{q\alpha}\right)^2+\left(\Gamma_{q\alpha}+\gamma_{k+q\mu}\right)^2/4}\right\},
\nonumber \\
&&\Lambda_{k\mu}^{M}(\omega)=\sum_{q\alpha}\left(f_{k\mu}^{q\alpha}\right)^2 \nonumber \\
&&\times \left\{n_{q\alpha}\ \frac{\Gamma_{q\alpha}+\gamma_{k+q\mu}}{\left(\omega-\omega_{k+q\mu}+\Omega_{q\alpha}\right)^2+\left(\Gamma_{q\alpha}+\gamma_{k+q\mu}\right)^2/4}\right. \nonumber
\\
&&+
\left.\left(1+n_{q\alpha}\right)\frac{\Gamma_{q\alpha}+\gamma_{k+q\mu}}{\left(\omega-\omega_{k+q\mu}-\Omega_{q\alpha}\right)^2+\left(\Gamma_{q\alpha}+\gamma_{k+q\mu}\right)^2/4}\right\}.\nonumber \\
\end{eqnarray}
For the phonons we have
\begin{eqnarray}
\Delta_{q\alpha}^{phon}(\omega)&=&\sum_{k\mu}\left(f_{k\mu}^{-q\alpha}\right)^2\left(N_{k-q\mu}-
N_{k\mu}\right)\nonumber \\
&\times &\frac{\omega-\omega_{k\mu}+\omega_{k-q\mu}}{\left(\omega-\omega_{k\mu}+\omega_{k-q\mu}\right)^2+\left(\gamma_{k\mu}+\gamma_{k-q\mu}\right)^2/4},
\nonumber \\
\Lambda_{q\alpha}^{phon}(\omega)&=&\sum_{k\mu}\left(f_{k\mu}^{-q\alpha}\right)^2\left(N_{k-q\mu}- N_{k\mu}\right)\nonumber \\
&\times &\frac{\gamma_{k\mu}+\gamma_{k-q\mu}}{\left(\omega-\omega_{k\mu}+\omega_{k-q\mu}\right)^2+\left(\gamma_{k\mu}+\gamma_{k-q\mu}\right)^2/4}.\nonumber \\
\end{eqnarray}
The steady state average number of photons and phonons are as in equations (\ref{SSNn}).

The many-particle effect for interacting photons and phonons in nanoscale structures manifests in the appearance of the effective damping rates and the renormalization frequency shifts. The main feature of the many-particle phenomena is the dependence of the effective damping rates and frequency shifts on the average number of photons and phonons within the waveguide. In the case of low density of photons and phonons and in the limit of weak photon-phonon coupling we get the independent photon and phonon SFs, the ones that are usually used for conventional quantum optomechanics, which are given by
\begin{eqnarray}\label{Empty}
\tilde{\cal
  S}_{k\mu}^{phot}(\omega)&=&\frac{\gamma_{k\mu}}{\left(\omega-\omega_{k\mu}\right)^2+\left(\gamma_{k\mu}\right)^2/4},\nonumber
\\
\tilde{\cal S}_{q\alpha}^{phon}(\omega)&=&\frac{\Gamma_{q\alpha}}{\left(\omega-\Omega_{q\alpha}\right)^2+\left(\Gamma_{q\alpha}\right)^2/4},
\end{eqnarray}
which are the usual Lorentzians.

\section{SFs of photon fields}

In this section we consider specific cases in which none, one or two photon modes of the waveguide to be excited. First, in order to emphasize the effect of thermal phonons, we consider a waveguide without excited photons. Second, we treat the case of a single photon mode at a fixed wavenumber of the lowest branch to be excited with a given average number of photons. Third, we discuss the case of two photon modes to be excited with a fixed average number of photons, and we concentrate in the possibility of heating or cooling of the phonon modes via the photon-phonon coupling.

In the present section we present the results for the case of photons from a single photon branch, hence we drop the photon branch index, $\mu$. Moreover, we assume the phonons to be in thermal equilibrium at temperature $T$, where the average number of phonons is given by
\begin{equation}
n_{q\alpha}=\frac{1}{e^{\frac{\hbar\Omega_{q\alpha}}{k_BT}}-1}.
\end{equation}
The thermal excitation of the optical photons is negligible at room temperature, and the photons can be excited only by an external pump. One can excite specific photon modes in order to get a fixed average number of photons in such modes, $N_{k}$. This result can be achieved via a combination of external pumps and dissipations. For simplicity, in the following we assume that the photon-phonon coupling parameter, and the photon and phonon damping rates, to be wavenumber independent in the appropriate zone of our interest. Namely, we assume $\gamma_k=\gamma$, $\Gamma_{q\alpha}=\Gamma_{\alpha}$ and $f_k^{q\alpha}=f_{\alpha}$.

\subsection{A cavity empty of photons}

We consider now the simple case of an empty waveguide, in which no photons appear inside the waveguide, but only thermal phonons exist. For optical photons one can neglect the photon thermal excitations. For empty waveguide we have
$N_k\approx 0$, and hence
$\Delta_{q\alpha}^{phon}(\omega)=\Lambda_{q\alpha}^{phon}(\omega)\approx 0$, with
$\Delta_{k}^{EM}(\omega)=\Lambda_{k}^{EM}(\omega)\approx 0$. For the
phonons we get the simple SF, $\tilde{\cal S}_{q\alpha}^{phon}(\omega)$, of equation (\ref{Empty}). The photon SF has contributions only due to the existence of thermal
phonons. We get ${\cal  S}_{k}^{phot}(\omega)$ of equation (\ref{SFPP}) with $\Delta_{k}^{phot}(\omega)=\Delta_{k}^{M}(\omega)$ and $\Lambda_{k}^{phot}(\omega)=\Lambda_{k}^{M}(\omega)$, of Eqs. (\ref{DLM}), in dropping the photon branch index $\mu$.

We concentrate now in the linear response of the system to a weak probe field of frequency $\omega$ and wavenumber $k$, when the cavity is empty of photons and only includes thermal phonons. The results are presented through the photon SF and its renormalization frequency shift and effective damping rate that are derived above. We present the results in using some physical numbers typical for nanoscale waveguides \cite{Zoubi2016}, (see discussion in section 2). For the photon and phonon dispersions we adopt the results in figures (2) and (3). The photon-phonon
coupling is taken to be $f_a/(2\pi)=f_v/(2\pi)\approx 1$~MHz. The photon damping rate is
$\gamma/(2\pi)\approx0.1$~MHz, the acoustic phonon damping rate is $\Gamma_a/(2\pi)\approx10$~MHz, and
the vibrational mode damping rate is $\Gamma_v/(2\pi)\approx1$~MHz. We consider
a temperature of $T=4$~K with average number of thermal vibration modes $n_v\approx 8$. In figure (4) we plot the photon frequency renormalization shift, $\Delta_{k}^{M}(\omega)$, as a function of the frequency $(\omega-\omega_{k+q})$, and in figure (5) we plot the effective photon damping rate, $\Lambda_{k}^{M}(\omega)$, as a function of the frequency $(\omega-\omega_{k+q})$. The photon SF, ${\cal
  S}_{k}^{phot}(\omega)$, is plotted in figure (6) as a function of $(\omega-\omega_{k})$. In the three plots we choose $ka=2$. The peaks appear at resonance for emission and absorption of phonon by photons subjected to conservation of energy and momentum. Therefore, the SFs allow the observation of the excited thermal phonons within the waveguide.
  
The effective damping rates of figure (5) and the frequency shifts of figure (4) correspond to the peaks of the SF of figure (6). Namely, the width of each peak in figure (6) of the SF is given by the effective damping rate of figure (5). The line width of the first peak is larger than the conventional damping rate. The width of the second peak is of the order of the conventional damping rate, and the rest peaks have smaller widths. The renormalization frequency shifts of the lines in figure (6) of the SF are given in figure (4), where the shifts can be positive or negative.

\begin{figure}
\includegraphics[width=0.8\linewidth]{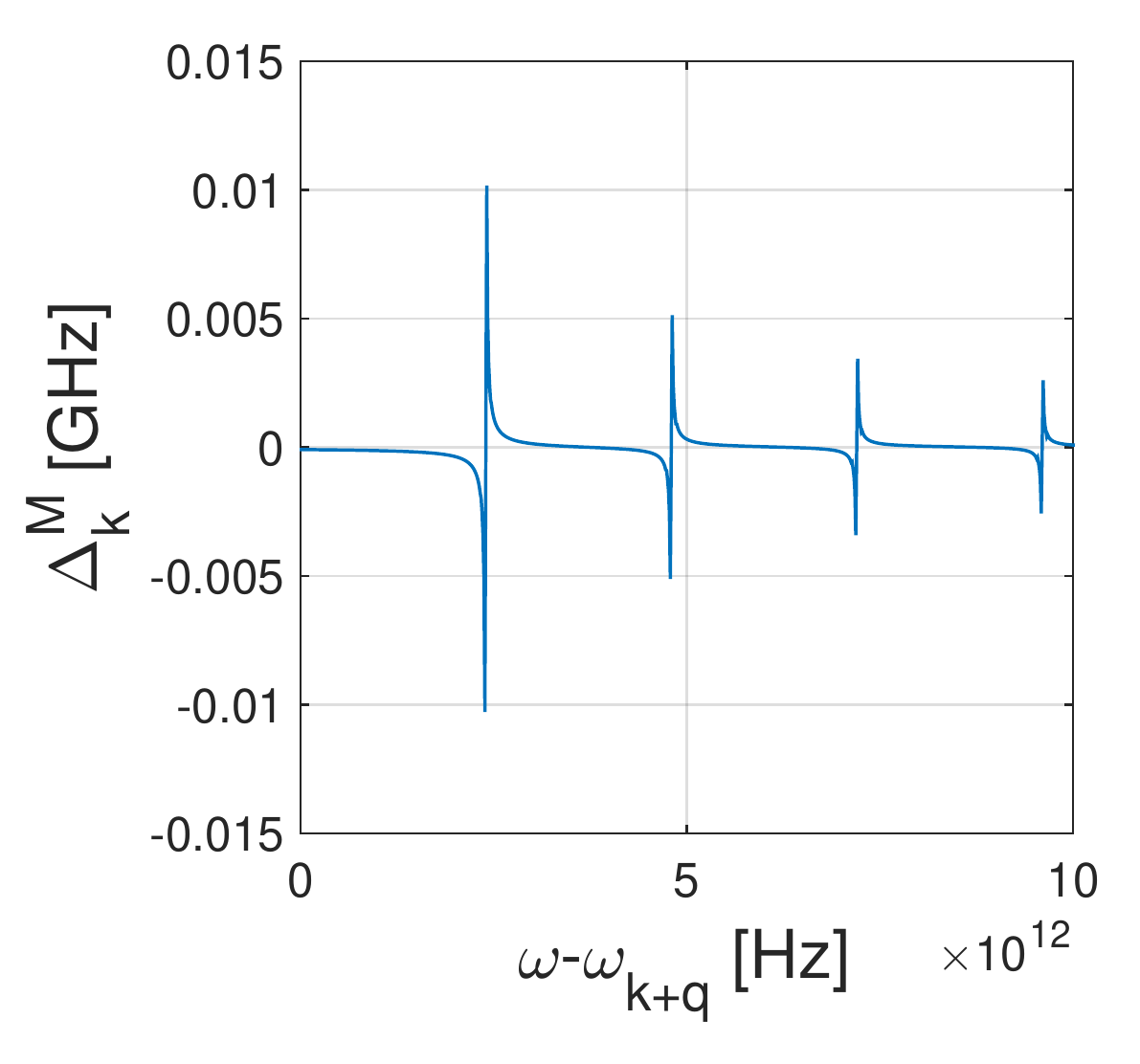}
\caption{The photon frequency shift, $\Delta_{k}^{M}(\omega)$, vs. $(\omega-\omega_{k+q})$, for $ka=2$.}
\label{PhotPhonDis}
\end{figure}

\begin{figure}
\includegraphics[width=0.8\linewidth]{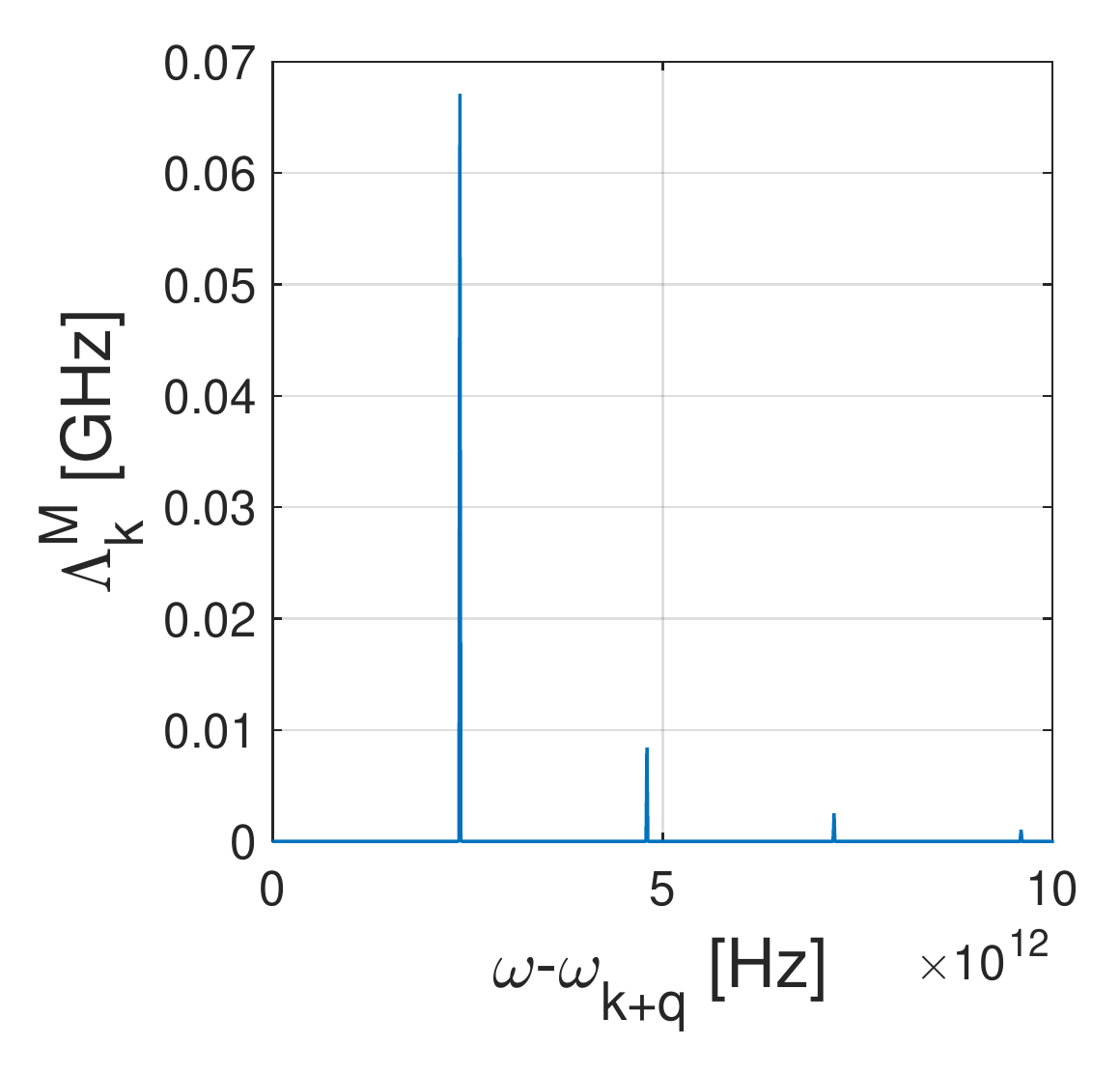}
\caption{The photon effective damping rate, $\Lambda_{k}^{M}(\omega)$, vs. $(\omega-\omega_{k+q})$, for $ka=2$.}
\label{PhotPhonDis}
\end{figure}

\begin{figure}
\includegraphics[width=0.8\linewidth]{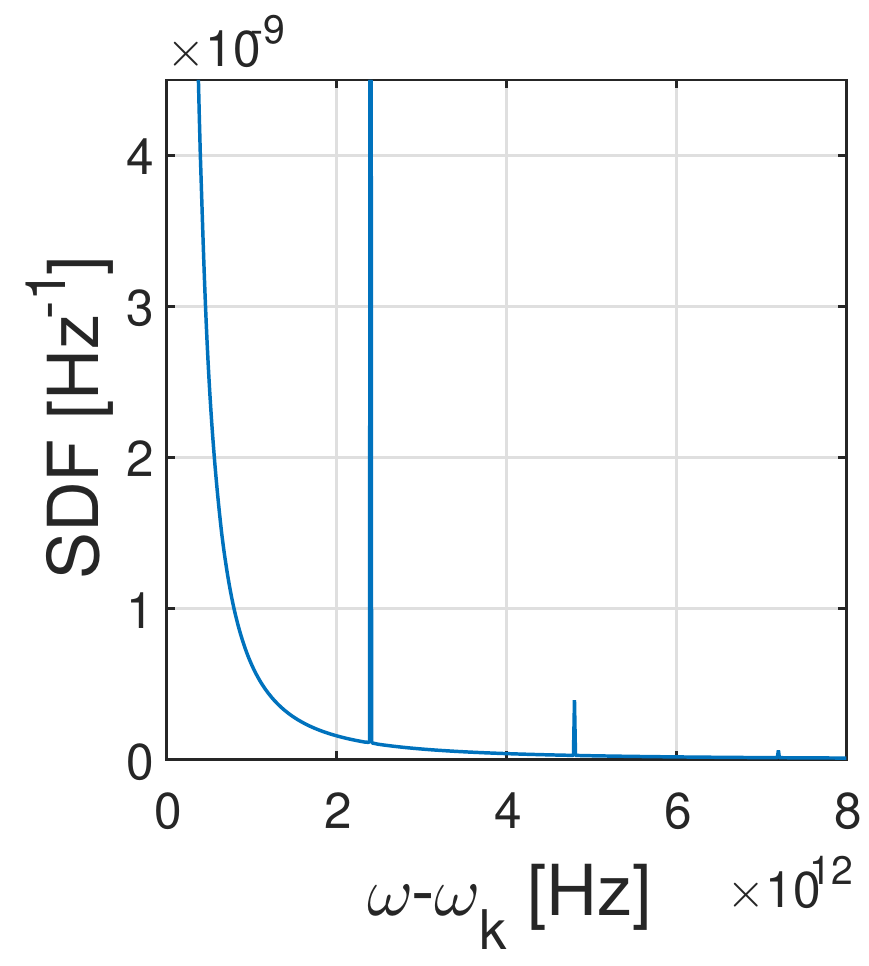}
\caption{The photon SF, ${\cal
  S}_{k}^{phot}(\omega)$, vs. $(\omega-\omega_{k})$, for $ka=2$.}
\label{PhotPhonDis}
\end{figure}

\subsection{A single photon field}

We assume a single mode that is excited inside the waveguide at wavenumber $k_0$. One can use an external probe field in order to excite the specific waveguide mode to get a fixed average number of photons with
$N_0$. Usually in nanoscale waveguides one has the limit of $\Gamma_{\alpha}\gg\gamma$. Then, in this case we get
\begin{eqnarray}
\Delta_{k}^{EM}(\omega)&=&\sum_{\alpha}f_{\alpha}^2N_{0}\ \left\{\frac{\omega-\omega_{k_0}-\Omega_{k_0-k}^{\alpha}}{\left(\omega-\omega_{k_0}-\Omega_{k_0-k}^{\alpha}\right)^2+\Gamma_{\alpha}^2/4}\right. \nonumber
\\
&-&
\left.\frac{\omega-\omega_{k_0}+\Omega_{k_0-k}^{\alpha}}{\left(\omega-\omega_{k_0}+\Omega_{k_0-k}^{\alpha}\right)^2+\Gamma_{\alpha}^2/4}\right\},
\nonumber \\
\Lambda_{k}^{EM}(\omega)&=&\sum_{\alpha}f_{\alpha}^2N_{0}\ \left\{\frac{\Gamma_{\alpha}}{\left(\omega-\omega_{k_0}-\Omega_{k_0-k}^{\alpha}\right)^2+\Gamma_{\alpha}^2/4}\right. \nonumber
\\
&-&
\left.\frac{\Gamma_{\alpha}}{\left(\omega-\omega_{k_0}+\Omega^{\alpha}_{k_0-k}\right)^2+\Gamma_{\alpha}^2/4}\right\},
\end{eqnarray}
For moderate excitation of the photons, up to one Watt power, one can neglect the effect of the thermal phonons on the photons. Hence, we neglect $\Lambda_{k}^{M}(\omega)$ and $\Delta_{k}^{M}(\omega)$. For the phonons we get
\begin{eqnarray}
\Delta_{q\alpha}^{phon}(\omega)&=&f_{\alpha}^2N_0\ \left\{\frac{\omega-\omega_{k_0+q}+\omega_{k_0}}{\left(\omega-\omega_{k_0+q}+\omega_{k_0}\right)^2+\gamma^2}\right. \nonumber \\
&-&\left. \frac{\omega-\omega_{k_0}+\omega_{k_0-q}}{\left(\omega-\omega_{k_0}+\omega_{k_0-q}\right)^2+\gamma^2}\right\},
\nonumber \\
\Lambda_{q\alpha}^{phon}(\omega)&=&f_{\alpha}^2N_0\ \left\{\frac{2\gamma}{\left(\omega-\omega_{k_0+q}+\omega_{k_0}\right)^2+\gamma^2}\right. \nonumber \\
&-&\left. \frac{2\gamma}{\left(\omega-\omega_{k_0}+\omega_{k_0-q}\right)^2+\gamma^2}\right\}.
\end{eqnarray}

We present the results using the previous typical numbers of a cylindrical nanoscale waveguide. We take a probe field at $k_0a=2$, and of power $1$~mW, that is $10^{16}$ photons per
second, with average number of photons in the waveguide of
$N_0\approx 10^6$. In figure (7) we plot the photon frequency shift, $\Delta_{k}^{EM}(\omega)$, as a function of the frequency $(\omega-\omega_{k+q})$, and in figure (8) we plot the effective photon damping rate, $\Lambda_{k}^{EM}(\omega)$, as a function of the frequency $(\omega-\omega_{k+q})$. In both plots we choose $k=k_0$, and the two resonances are only due to the vibrational mode. In the right side of figure (8) the effective damping is positive as a field photon jump to a lower photon mode by the emission of a vibrational mode, while the left side is negative as a higher photon mode jump into the field photon by the emission of a vibrational mode. The effective damping rate is much larger than the conventional damping rate, that is we get $\Lambda_{k}^{EM}(\omega)
\gg\Gamma_{\alpha}$, and we have a significant frequency shift $\Delta_{k}^{EM}$.

We treated the effect of thermal phonons in the previous case of empty cavity, and here we considered a single excited field in neglecting the influence of thermal phonons. Including the effect of thermal phonons to the present case of a single excited field will lead to the total photon frequency shift $\Delta_{k\mu}^{phot}(\omega)=\Delta_{k\mu}^{M}(\omega)+\Delta_{k\mu}^{EM}(\omega)$ and to the total effective photon damping rate
$\Lambda_{k\mu}^{phot}(\omega)=\Lambda_{k\mu}^{M}(\omega)+\Lambda_{k\mu}^{EM}(\omega)$. Now, the average number of photons in the specific mode is much larger than the average number of thermal phonons, that is $N_0\gg n$. Adding the effect of thermal phonons, that are plotted in figures (4) and (5), will appear as additional background noise in figures (7) and (8). It is clear, by comparison between the figures, that in the present case of high intensity field the effect of thermal phonons is negligible.

\begin{figure}
\includegraphics[width=0.8\linewidth]{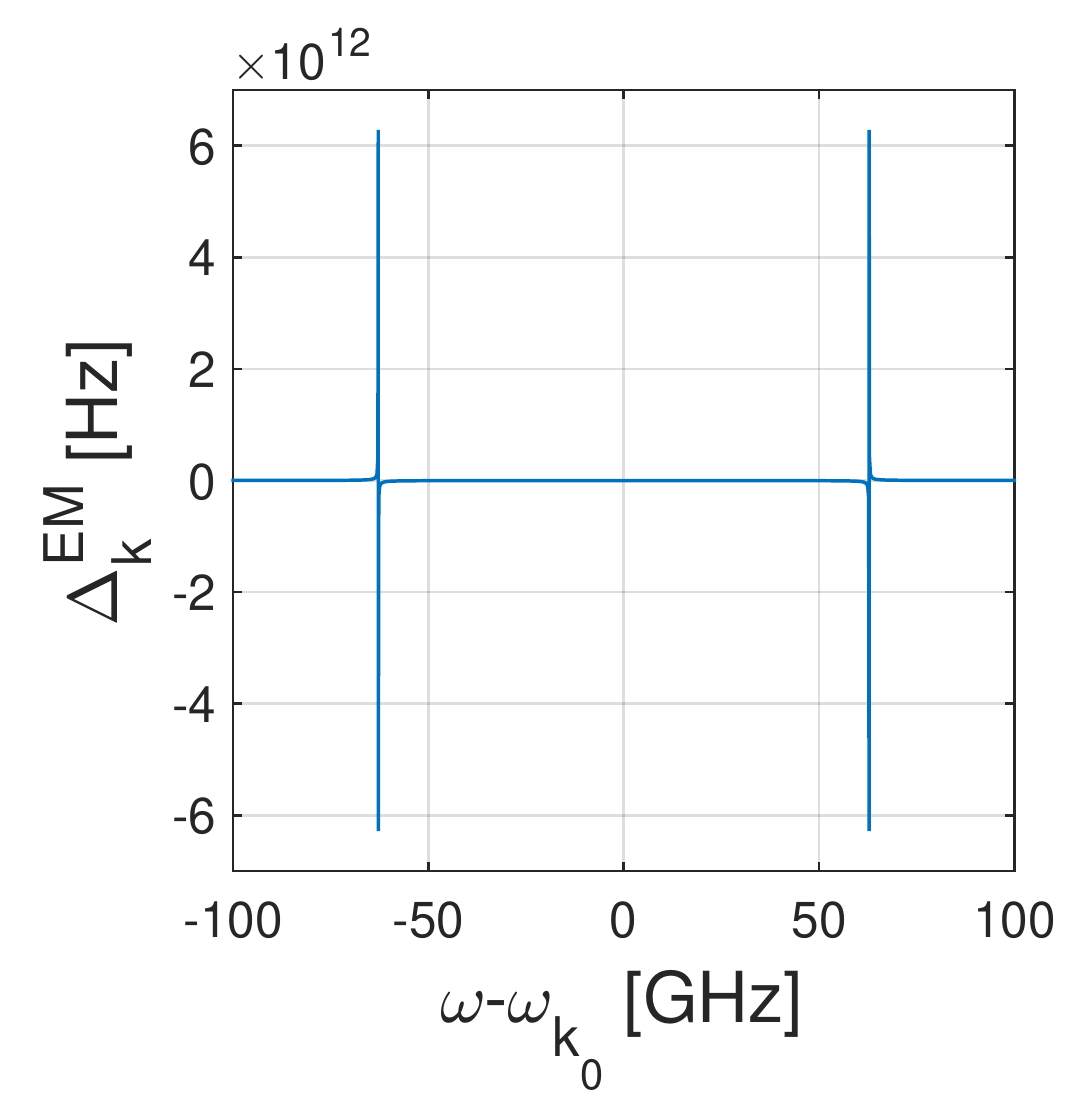}
\caption{The photon frequency shift, $\Delta_{k}^{EM}(\omega)$, vs. $(\omega-\omega_{k_0})$, for $ka=k_0a=2$.}
\label{PhotPhonDis}
\end{figure}

\begin{figure}
\includegraphics[width=0.8\linewidth]{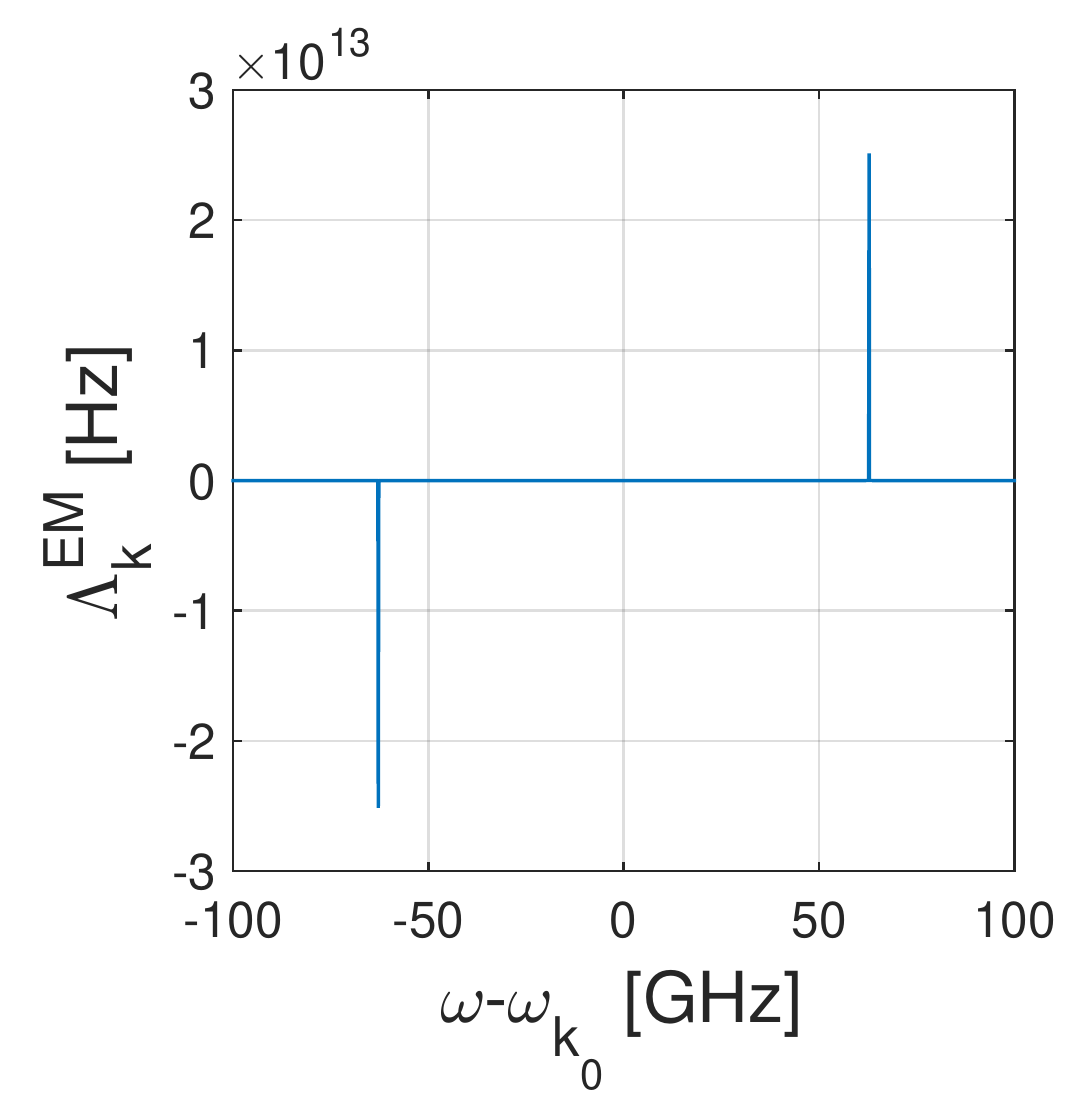}
\caption{The photon effective damping rate, $\Lambda_{k}^{EM}(\omega)$, vs. $(\omega-\omega_{k_0})$, for $ka=k_0a=2$.}
\label{PhotPhonDis}
\end{figure}

\subsection{Two photon fields}

We consider next the case of two waveguide modes to be exited at $k_1$ and $k_2$ using external fields. The average number of waveguide photons at the two specific modes are $N_1$ and
$N_2$. For this case we obtain
\begin{eqnarray}
\Delta_{k}^{EM}(\omega)&=&\sum_{i\alpha}f_{\alpha}^2N_{i}\ \left\{\frac{\omega-\omega_{k_i}-\Omega^{\alpha}_{k_i-k}}{\left(\omega-\omega_{k_i}-\Omega^{\alpha}_{k_i-k}\right)^2+\Gamma_{\alpha}^2/4}\right. \nonumber
\\
&-&
\left.\frac{\omega-\omega_{k_i}+\Omega^{\alpha}_{k_i-k}}{\left(\omega-\omega_{k_i}+\Omega^{\alpha}_{k_i-k}\right)^2+\Gamma_{\alpha}^2/4}\right\},
\nonumber \\
\Lambda_{k}^{EM}(\omega)&=&\sum_{i\alpha}f_{\alpha}^2N_{i}\ \left\{\frac{\Gamma_{\alpha}}{\left(\omega-\omega_{k_i}-\Omega^{\alpha}_{k_i-k}\right)^2+\Gamma_{\alpha}^2/4}\right. \nonumber
\\
&-&
\left.\frac{\Gamma_{\alpha}}{\left(\omega-\omega_{k_i}+\Omega^{\alpha}_{k_i-k}\right)^2+\Gamma_{\alpha}^2/4}\right\},
\end{eqnarray}
in the limit of $\Gamma_{\alpha}\gg\gamma$, and as before we neglect $\Lambda_{k}^{M}(\omega)$ and $\Delta_{k}^{M}(\omega)$. For the phonons we get
\begin{eqnarray}
\Delta_{q\alpha}^{phon}(\omega)&=&\sum_if_{\alpha}^2N_i\ \left\{\frac{\omega-\omega_{k_i+q}+\omega_{k_i}}{\left(\omega-\omega_{k_i+q}+\omega_{k_i}\right)^2+\gamma^2}\right. \nonumber \\
&-&\left. \frac{\omega-\omega_{k_i}+\omega_{k_i-q}}{\left(\omega-\omega_{k_i}+\omega_{k_i-q}\right)^2+\gamma^2}\right\},
\nonumber \\
\Lambda_{q\alpha}^{phon}(\omega)&=&\sum_if_{\alpha}^2N_i\ \left\{\frac{2\gamma}{\left(\omega-\omega_{k_i+q}+\omega_{k_i}\right)^2+\gamma^2}\right. \nonumber \\
&-&\left. \frac{2\gamma}{\left(\omega-\omega_{k_i}+\omega_{k_i-q}\right)^2+\gamma^2}\right\}.
\end{eqnarray}
At resonance, where the difference between the two fields fits exactly with one
of the phonons, that is $k_1-k_2=q$ and $\omega_1-\omega_2=\Omega^{\alpha}_{q_0}$,
then one channel becomes dominant and contributes
\begin{eqnarray}
\Delta_{q_0\alpha}^{phon}(\omega)&=&f_{\alpha}^2\left(N_2-N_1\right)\ \frac{\omega-\Omega^{\alpha}_{q_0}}{\left(\omega-\Omega^{\alpha}_{q_0}\right)^2+\gamma^2},
\nonumber \\
\Lambda_{q_0\alpha}^{phon}(\omega)&=&f_{\alpha}^2\left(N_2-N_1\right)\ \frac{2\gamma}{\left(\omega-\Omega^{\alpha}_{q_0}\right)^2+\gamma^2}.
\end{eqnarray}
The sign of $N_2-N_1$ is of big importance and fixes if we have heating or cooling of the phonon mode
$(q_0)$ \cite{Otterstrom2018}. As $\omega_1>\omega_2$ then for $N_2>N_1$ we get cooling, and for $N_1>N_2$ we get heating. At resonance the damping rate is $\Lambda_{q_0\alpha}^{phon}=\frac{2f_{\alpha}^2}{\gamma}\left(N_2-N_1\right)$. Note that one can easily achieve $\Lambda_{q_0\alpha}\gg\Gamma_{\alpha}$ by taking the limit $N_1\ll N_2$.

\section{Conclusions}

Green's functions and SFs are widely used tools that provide physical properties of many-particle problems. Here we applied the technique for interacting photons and phonons in extended nanoscale structures, where we treated a one dimensional waveguide made of high contrast dielectric material that embedded in free space. In such devices, photons and phonons can freely propagate along the waveguide axis with wavenumbers and interact with a Brillouin scattering type Hamiltonian. The nonlinear Hamiltonian includes three-particle processes in which a photon enters and another leaves with the emission or absorption of a phonon. The photon and phonon SFs are derived from the retarded photon and phonon Green's functions, respectively. In the spirit of response theory, for these functions a test particle is created at a point inside the waveguide and annihilated at another point. We work in Fourier space and use wavenumbers as good quantum numbers, which is justified due to translational symmetry along the waveguide axis. The real space description is straightforward achievable in applying Fourier transform with a given weight function.

We solved for the Green's functions under the factorization approximation of the mean-field theory, in which photon-phonon correlations are neglected due to fast dephasing in the setup. The results hold for the general case of photons and phonons of several branches, while we concentrated in the case of lower branch photons and the lowest two phonon branches of acoustic waves and vibrational modes. The calculations yield effective damping rates for both photons and phonons due to many-particle phenomena, and which lead to broadening that are related on the average number of excitations within the waveguide. Note that such broadening appear in addition to the conventional phonon and photon damping rates, that are included phenomenologically in the paper. Moreover, we obtain renormalized frequencies for both the photons and the phonons, that lead to energy shifts that depend on density of excitations. We studied several interesting cases, for example we emphasized the effect of thermal phonons on the scattering of photons. We treated cases where one or two photon modes to be highly excited with a fixed average number of photons, which can be achieved by a combination of external pump fields and leaks of photons out of the waveguide, that can be handled using input-output formalism. The case of two excited modes show phonon heating and cooling phenomena that can be exploited for phonon cooling scenarios.

In extended nanoscale waveguides the multi-mode nature of photons and phonons implies treating the system as a many-particle problem. Therefore, it is necessary to extend the conventional quantum optomechanics into continuum quantum optomechanics. Nanoscale waveguides are promising candidates for the physical implementation of quantum information processing, and hence coherent behavior of photons and phonons in such setups is critical for efficient manipulation and performance. Then the simple SFs, which are derived for single photon and phonon modes of quantum optomechanics, are insufficient for continuum optomechanics, mainly due many-particle phenomena in extended systems. In the present paper we concentrated in the case of steady state at thermal equilibrium, and we plan to extend our study in the future into non-equilibrium state.

\section*{Acknowledgment}

The work was supported by the Council for Higher Education in Israel via the Maa'of Fellowship.

\end{document}